\documentclass[prd,twocolumn,preprintnumbers,amsmath,amssymb,nofootinbib]{revtex4-1}
\usepackage{mathrsfs,}
\usepackage{color}
\usepackage[active]{srcltx}
\usepackage{amsmath,amsfonts,amssymb,amsthm,amstext,amscd,eucal,srcltx}
\usepackage{epsfig,graphicx,bm}
\usepackage{epstopdf, epsf}
\usepackage{dcolumn}
\usepackage{hyperref}

\def\nn{\nonumber}

\newcommand{\mpl}{M_{\rm pl}}

\def\AS{{\alpha_{k}^*}}
\def\Ai{{\alpha_{k_i}}}

\def\Ais{{\alpha_{ k_i}^{*}}}
\def\Ajs{{\alpha_{ k_j}^{*}}}
\def\A{{\alpha_{k}}}
 \def\B{{\beta_{k}}}
\def\Bi{{\beta_{ k_i}}}

\def\Bis{{\beta_{k_i}^{*}}}
\def\Bjs{{\beta_{k_j}^{*}}}
 \def\BS{{\beta_{k}^*}}

\def\chis{\chi_{{}_S}}

\def\phis{\varphi_{{}_S}}
\def\vare{{\varepsilon}}
\def\b1{{\beta_1}}
\def\cospsi{{\cos\psi_{\hat x}}}
\def\cossq{{{\cos}^2 \psi_{\hat x}}}
\def\nn{\nonumber \\ }

 \newcommand{\be}{\begin{equation}}
\newcommand{\ee}{\end{equation}}
\newcommand{\beqa}{\begin{eqnarray}}
\newcommand{\eeqa}{\end{eqnarray}}

\begin{document}
\title{Hemispherical Anomaly from Asymmetric Initial States}
\author{Amjad Ashoorioon$^1$, Tomi Koivisto$^2$ }
\bigskip\medskip
\affiliation{$^1$~ Institutionen f\"{o}r fysik och astronomi
Uppsala Universitet, Box 803, SE-751 08 Uppsala, Sweden\\
$^2$ Nordita, KTH Royal Institute of Technology and Stockholm University,
Roslagstullsbacken 23, SE-10691 Stockholm, Sweden\\ }
\vfil
\preprint{NORDITA-2015-88} 
\preprint{UUITP-14/15} 
\begin{abstract}

We investigate if  the hemispherical asymmetry in the CMB is produced from ``asymmetric'' excited initial condition. We show that in the limit where the deviations from the Bunch-Davies vacuum is large and the scale of new physics is maximally separated from the inflationary Hubble parameter, the primordial power spectrum is modulated only by position dependent dipole and quadrupole terms. Requiring the dipole contribution in the power spectrum to account for the observed power asymmetry, $A=0.07\pm0.022$, we show that the amount of quadrupole terms is roughly equal to $A^2$. The {\it mean} local bispectrum, which gets enhanced for the excited initial state, is within the $1\sigma$ bound of Planck 2015 results for a large field model, $f_{\rm NL}\simeq 4.17$, but is reachable by future CMB experiments.  The amplitude of the local non-gaussianity modulates around this mean value, depending on the angle that the correlated patches on the  2d CMB surface make with the preferred direction. The amount of variation minimizes for the configuration in which the short and long wavelengths modes are  around the preferred pole and $|\vec k_3|\approx |\vec k_{l\approx10}|\ll |\vec k_1|\approx |\vec k_2|\approx |\vec k_{l\approx2500}|$ with $f_{\rm NL}^{\rm min}\approx 3.64 $. The maximum occurs when these modes are at the antipode of the preferred pole, $f_{\rm NL}^{\rm max}\approx 4.81$ . The difference of non-gaussianity between these two configurations is as large as $\simeq 1.17$ which can be used to distinguish this scenario from other scenarios that try to explain the observed hemispherical asymmetry.  \end{abstract}

\maketitle

\section{Introduction}

Inflation, despite successfully explaining the general pattern of observed cosmic microwave background radiation (CMB) \cite{Ade:2015lrj}, fails to explain few anomalies at large scales. Some of these anomalies, which were previously observed in the WMAP data \cite{Eriksen:2003db}, persist even in the latest  Planck data, even though their statistical significances might not be substantial \cite{Ade:2015oya}. Such anomalies in general break the statistical isotropy of the CMB and can be modeled phenomenologically as
\be\label{DeltaT}
\Delta T(\hat x)=  \Delta T_{\rm iso}(\hat x) (1+M(\hat x))\,,
\ee
where $T_{\rm iso}$ is the isotropic part of the temperature fluctuations, where $\hat x$ is where you look in the sky. In particular, there seems to be hemispherical asymmetry consistent with the existence of a dipolar modulation term, $M(\hat x)=A \hat{x}\cdot\hat n$ in the Planck data  with amplitude $A\approx 6-7\%$
 on scales $2\leq l\leq 64$ \cite{Ade:2015lrj,aiola, Ade:2015oya}.  As stated before, $\hat{x}$ is where you look in the sky and $\hat n$ is the preferred direction. The asymmetry seems to fade away at smaller scales, especially for $l\geq 600$ \cite{Flender:2013jja,  Ade:2015hxq}. The asymmetry is more than twice as large as the expected asymmetry due to cosmic variance, $A\approx 2.9\%$. Similar asymmetry can arise from a dipolar term in the primordial power spectrum \cite{Koivisto:2010fk} or from a phenomenological $x$-dependent modulation of the primordial spectrum \cite{pv-spectrum}
\be\label{P-parity}
{\mathcal P}_{S}= {\mathcal P}_{\rm iso} \left(1+ 2A (\hat{x}\cdot\hat{n}) \right)\,.
\ee
Various proposals have been offered to explain this asymmetry. Some considered the effect of long wave-length super-horizon mode on the sub-horizon power spectrum \cite{Erickcek:2008sm,pv-spectrum} through Grishchuk-Zel'dovick effect \cite{GZ-effect}, which requires non-negligible local non-gaussianity to correlate the long and short wavelength modes. Also noncommutative physics at Planck scale \cite{Koivisto:2010fk}, isocurvature perturbation \cite{Erickcek:2009at}, non-Gaussianity \cite{adhikari,Kenton:2015jga}, domain walls \cite{Jazayeri:2014nya} or running of scalar spectral index \cite{McDonald:2014kia} have been suggested as mechanisms  explaining the observed asymmetry in the power spectra. In principle, higher order multipoles can also contribute to  \eqref{DeltaT}, and respectively \eqref{P-parity}.
 
%
The main goal of this paper is to design a scenario in which the observed dipole asymmetry is realized, assuming that the initial condition for fluctuations has a small  anisotropic position-dependent asymmetric part\footnote{By ``asymmetric'', we mean that the initial condition for the scalar perturbations  is not invariant under the the transformation $\mathbf{x}\rightarrow -\mathbf{x}$ within the horizon patch (by horizon patch we mean the patch that becomes the size of the current observable universe after inflation and subbsequent stages of cosmological evolution.).}. Such asymmetric-contribution in the initial condition could be the effect of preinflationary patch which was probably highly inhomogeneous and anisotropic or the effect of parity violating terms in the fundamental theory higher than the energy scale of inflation, which was also position dependent within the inflationary patch. The quantum vacuum state of the universe is thus not assumed to initially be aligned with the Bunch-Davies vacuum which is the standard choice. In reference \cite{Schmidt:2012ky} non-Bunch Davies vacuum was also considered as a possible reason for the power asymmetry, but their mechanism, that was based on coupling of modes in an isotropic vacuum, was different.

 Fixing the amount of the asymmetry from the observed hemispherical asymmetry, we  also predict a non-negligible quadrupole contribution to the primordial power spectrum 
 \be\label{P-modulated}
{\mathcal P}_{S}= {\mathcal P}_{\rm iso} \left(1+ 2A (\hat{x}\cdot\hat{n}) +B(\hat{x}\cdot\hat{n}) ^2 \right)\,.
\ee
 with $B$ within the interval\footnote{In models of anisotropic inflation \cite{male}, generally quadrupole term, ${(\hat{k}\cdot\hat{n})}^2$ in momentum space appears which is different from the quadrupole term in position space we predict here.}
\be
0.0025\leq B\simeq A^2 \leq 0.008\,.
\ee
In our scenario, higher order multipole contributions to the primordial power spectrum will not only be suppressed by higher powers of $A$, which is small, but also by negative powers of $N_k\gg1$, where $N_k$ is the number of quanta in the initial excited state which is related to the second Bogoliubov coefficient $\beta_k$ through the relation $N_k\equiv |\beta_k|^2$. As shown in \cite{Ashoorioon:2013eia}, one can start from excited initial states with large occupation number,  $N_k\gg1$ without violating the bounds on backreaction. 

The structure of the paper is as follows. First we review the formalism of excited initial condition, showing how dipole and quadrupole terms could be generated from a asymmetric excited initial condition. As expected, the local configuration is enhanced for such excited initial states with an amplitude which is within the $1\sigma$ bound of Planck data. However, in addition, one finds an angular-dependent modulation that depends on the direction that each mode is located at and the angle that it makes with the preferred direction. The amplitude of  modulation minimizes for the local configuration around the preferred pole, in which the short wavelength modes are the smallest one that could be probed by Planck, $l\approx2500$ and the longest one corresponds to $l\approx 10$ at which the cosmological variance is small. For the same local configuration which is at the antipode pole the local non-gaussianity reaches its maximum. We conclude the paper at the end.

\section{asymmetric Excited Initial States}

  As quite well-known, the predictions of inflationary models for the CMB spectrum depends on the initial state of the quantum perturbations as well as the specific details of the model. The standard lore is that these perturbations embark upon the Bunch-Davis (BD) vacuum \cite{Bunch:1978yq}, which is the minimum energy states, when they pop out of vacuum inside the horizon  of an inflationary background. However,
 various effects of physics at energy scales higher than that of inflation \cite{Initial-data-literature} or multi-field effects \cite{Shiu:2011qw} can excite these fluctuations to a state other than the Bunch-Davies vacuum.
In a previous work, we showed that how by assuming initial conditions  other than the Bunch-Davies vacuum, one can decrease the tensor/scalar ratio in a high energy scale chaotic models like $m^2\phi^2$ \cite{Ashoorioon:2013eia}  and make it compatible with the latest Planck data \cite{Ade:2015lrj,Ade:2013ydc}. We also showed how one can induce large amount of running in the scalar spectral index or blue tensor spectral index using scale-dependent initial condition \cite{Ashoorioon:2014nta}. It is shown that excited initial state can induce larger $\mu$-type distortions in comparison with the Bunch-Davies vacuum \cite{Ganc:2012ae}.

The equation of motion for the gauge-invariant scalar perturbations, the Mukhanov-Sasaki variable $u(\tau,y)$ \cite{Mukhanov:1990me},
\begin{equation}\label{u-mukhanov}
u=-z \left( \frac{a^{\prime}}{a}\frac{\delta \phi}{\phi^{\prime}}+\Psi\right), \quad z\equiv \frac{a \phi^{\prime}}{\cal H}, \quad {\cal H}\equiv \frac{a^{\prime}}{a},
\end{equation}
is
\begin{equation}\label{u-eq}
u^{\prime\prime}_{ k}+\left(k^2-\frac{z^{\prime\prime}}{z}\right)u_{k}=0\,.
\end{equation}
Prime denotes derivative with respect to the conformal time $\tau$ and $u_k(\tau)$ is the Fourier mode of $u(\tau,y)$.
For a quasi-de-Sitter background
 \begin{equation}\label{background}
 a(\tau)\simeq -\frac{1}{H\tau}\,,
 \end{equation}
where $H$ is the Hubble constant. The most generic solution to \eqref{u-eq} with (\ref{background}) is 
\begin{equation}\label{u-sol-ds}
u_{k}(\eta)\simeq\frac{\sqrt{\pi|\tau|}}{2}\left[\A H_{3/2}^{(1)}(k|\tau|)+\B H_{3/2}^{(2)}(k|\tau|)\right]\,,
\end{equation}
where $H_{3/2}^{(1)}$ and $H_{3/2}^{(2)}$ are respectively  Hankel functions of the first and second kind. The terms proportional to $\AS$ and $\BS$ respectively behave like the positive and negative frequency modes in infinite past. These Bogoliubov coefficients satisfy the Wronskian constraint,
\begin{equation}\label{Wronskian}
|\A|^2-|\B|^2=1\,.
\end{equation}
The standard BD vacuum is obtained when $\A=1$ and $\B=0$. 

For a generic initial state,  the energy and pressure density carried by the fluctuations are of the same order, $\delta p_{\text{non-BD}} \sim \delta \rho_{\text{non-BD}}$,  and should remain subdominant with respect to the total energy of the inflaton. Also their variations with time should not hinder the slow-roll inflation. Noting that $\delta\rho_{\text{non-BD}}'\sim \delta p_{\text{non-BD}}'\sim {\cal H} \delta\rho_{\text{non-BD}}$ in the leading slow-roll approximation, this requirement is satisfied if
\be\label{background-backreaction}
\delta\rho_{\text{non-BD}}\ll \epsilon\rho_0\,,\quad \delta p_{\text{non-BD}}'\ll {\cal H}\eta\epsilon\rho_0\,,
\ee
where $\epsilon$ and $\eta$ are defined as
 \begin{eqnarray}\label{slow-roll-def}
\epsilon\equiv 1-\frac{{\cal H}^{\prime}}{{\cal H}^2}\ll 1\,,\qquad \eta\equiv\epsilon-\frac{\epsilon^{\prime}}{2{\cal H}\epsilon}\ll 1\,.
\end{eqnarray}
The strongest of the above two constraints may be written in terms of $\beta_k$ as
\be\label{energy-excited}
\int_H^{\infty} \frac{{\rm d}^3 k}{(2\pi)^3} k |\B|^2\ll \epsilon\eta H^2\mpl^2\,.
\ee
We will assume that all scales of interest are uniformly excited to an initial state with the second Bogolibubov coefficient, which is anisotropic in position space within the initial inflating patch,
\be\label{beta0}
\beta_{k}=\beta_0(\hat x)\,,
\ee
once their physical momenta become smaller than the scale of new physics, $M$ \cite{danielsson}.  Inevitably, modes which remain above this hypersurface momentum do not get excited and therefore the left hand side of the integral \eqref{energy-excited} remains finite\footnote{There is a qualitatively different situation in which at time $\tau_0$, the modes with physical momentum smaller than $M$ get pumped to an excited state, whereas the larger ones remain in their vacuum. These two pictures, even though are qualitatively different, lead to the same result quantitatively. For the latter to be relevant for the CMB scales, one would expect that inflation did not last more than what is needed to solve the problems of Big Bang cosmology.}.  With the choice \eqref{beta0}, one does not lead to extra $k$-dependence in the power spectra and does not change the spectral index at the observable scales. As mentioned, we also assume that the second Bogoliubov coefficient can depend on the position direction through the parameter $\beta_0(\hat x)$. Having
\begin{equation}\label{p-rho-massless-quanta}
\delta \rho_{\text{non-BD}} \sim |\beta_0(\hat x) |^2 M^4\,,\quad
\delta p_{\text{non-BD}}'/{\cal H}\sim |\beta_0(\hat x)|^2  M^4\,,
\end{equation}
one obtains the following upper bound on $|\beta_0(\hat x)|$,
\begin{eqnarray}\label{beta-scalar-backreaction}
 |\beta_0 (\hat x)| \lesssim \sqrt{\epsilon\eta}\frac{H M_{\rm Pl}}{M^2}\sim \epsilon\frac{H M_{\rm Pl}}{M^2}\,.
\end{eqnarray}
As it was discussed in \cite{Ashoorioon:2013eia} and will be reviewed briefly below, $|\beta_0(\hat x)|$ is not necessarily very small. Larger values of $|\beta_0 (\hat x)|$ are compensated with a smaller Hubble parameter, $H$, for a given model to match the normalization of density perturbations with the data.

The scalar power spectrum defined as,
\begin{equation}
{\mathcal P}_{S}=\frac{k^{3}}{2\pi ^{2}}\left| \frac{u_{k}}{z}\right|^2_{{k/{\cal H}\rightarrow 0}}\,,
\label{scrpower}
\end{equation}
turns out to be
\begin{equation}\label{power-spectrum-scalar}
{\mathcal P}_S={\mathcal P}_{BD}\,\gamma_{{}_S}\,,
\ee
where
\be\label{P-BD-gamma-S}
{\mathcal P}_{BD}=\frac{1}{8\pi^2\epsilon}\left(\frac{H}{\mpl}\right)^2,\qquad \gamma_{{}_S}=|\alpha_{k}^S-\beta_{k}^S|^2_{{}_{k={\cal H}}}\,.
\end{equation}

To study this power spectrum more closely, we note that the energy and the power spectra (and also the bi-spectrum) expressions only depend on relative phase of $\alpha,\ \beta$. Hence, they may be parameterized as
\be\label{parametrization}
\begin{split}
\alpha^S_{ k}=\cosh\chi_{{}_S} e^{i\varphi_{{}_S}}\,&,\quad \beta^S_{k} =\sinh\chi_{{}_S} e^{-i\varphi_{{}_S}}\,.\\
\end{split}
\ee
With this parametrization, $\chi_{{}_S}\simeq\sinh^{-1}\beta_0(\hat x),\ e^{-2\chi_{{}_S}}\leq \gamma_{{}_S}\leq e^{2\chi_{{}_S}}$. As it was shown in \cite{Ashoorioon:2013eia},  in the regime where the deviation from the Bunch-Davies vacuum is large, $\chis \gg 1$, in order to have maximal separation between the scale of new physics, $M$, and the inflationary Hubble parameter, $H$, one is confined to $\phis\simeq \pi /2$. For $m^2\phi^2$, for large $\chis \gg 1$, $M\simeq 21 H$.

Let us now assume that the horizon patch or the asymmetric effect of new physics at the energy scale higher than the energy scale of inflation is anisotropic and in particular it singles out one direction such that the  parameter $\beta_0(\hat x)$ in the initial state Bogoliubov coefficient involves an asymmetric term too. We parameterise this asymmetric effect at the new physics hypersurface with $\vare$ as follows\footnote{In this paper, we assume that the mechanism that excites the fluctuations within the horizon patch is position-dependent and has picked up a small dipole-dependent correction in addition to the usual monopole homogeneous term. This in particular is conceivable if one assumes that the horizon patch was bigger than a Hubble size and thus the mechanism responsible for the excitement of the mode leads to different values in different parts of the horizon patch.  In the first approximation we assumed that the second Bologoliubov coefficient has a small dipole correction in addition to the uniform part.  The anisotropic vacuum that we have hypothesised in our article to justify the hemispherical asymmetry would correspond to an anisotropic energy momentum tensor ``within'' the horizon patch. This should be compared with the situation described by Erickcek {\it et al.}  \cite{Erickcek:2008sm}, where a mode larger than our current horizon creates this asymmetry. In that scenario, it is not clear why a mono-wavelength superhorizon anisotropy in a particular direction should be hypothesised to obtain the observed hemispherical asymmetry.  The observation of the hemispherical asymmetry in the CMB in that scenario provides information about the distribution of energy momentum tensor at the scales beyond our horizon. Another variant of this scenario which includes all superhorizon modes and non-gaussianity on scales larger than the scale of our universe are considered \cite{adhikari}, tackles the fine-tuning issue better. As we will see, what our computations show is that if the energy-momentum tensor of the inflaton  is anisotropic ``within the horizon patch'' in the beginning of inflation, no matter how long the subsequent inflation lasts, the resulted power spectrum is anisotropic.}
\be
\beta_0(\hat x)=\sinh\chis (1+\vare \hat{x}.\hat{n}) e^{i\phis}\,,
\ee
where $\vare\lesssim 1$. We may interchangeably use $\cos\psi_{\hat{x}}$ for $\hat{x}\cdot\hat{n}$ in the rest of the analysis. From the Wronskian constraint \eqref{Wronskian}, one can easily obtain the norm of the first Bogoliubov coefficient too. Following the parameterization of \cite{Ashoorioon:2013eia}  the first Bogoliubov coefficient takes the form
\be
\AS=[1+\sinh \chis (1+\vare \cospsi)^2]^{1/2} e^{-i\phis}\,.
\ee
One can easily obtain the factor $\gamma_S$ through the relation \eqref{P-BD-gamma-S}. Expanding $\gamma_S$ as a function of $\vare$, one obtains
\beqa
\gamma_S&=& \gamma_0+\vare \cospsi \gamma_1+\vare^2  \cossq\gamma_2+\ldots\nonumber\\
                   &=&\cosh 2 \chis-\cos 2\phis \sinh 2\chis \nonumber\\
                  &+& \vare \cospsi \left[2 \tanh \chis (\sinh 2 \chis-\cos 2 \phis \cosh 2 \chis) \right]\nonumber\\
                  &+& \vare^2 \cossq  \left[ 2 \sinh ^2\chis-\cos 2 \phis (\cosh 2 \chis+2)\right.\nn&& \left. \tanh ^3\chis \right]+\ldots\nonumber\\
\eeqa
where $\gamma_i$'s are the level $i$-th order coefficient in $\vare$ expansion of $\gamma_S$ and the ellipses stands for the higher order of $\vare$ which are suppressed. The higher order terms are not only suppressed by powers of $\vare$, but also by powers of $e^{-2\chis}$ where $\chis\gg1$, which make them completely negligible in comparison with other terms. In this limit, the primordial scalar power spectrum obtains  dipole and quadrupole directional dependence
\be
{\mathcal P}_{S}= {\mathcal P}_{\rm iso} \left(1+ 2A (\hat{x}\cdot\hat{n})+B (\hat{x}\cdot\hat{n})^2 \right)\,.
\ee
The parameters $A$ and $B$ are respectively defined as
\beqa
A&\equiv& \frac{\vare \gamma_1}{2~\gamma_0}\,,\\
B &\equiv& \frac{\vare^2 \gamma_2}{\gamma_0}\,.
\eeqa
In the large $\chis$ limit, $\chis \gg 1$, $A$ and $B$ can be expanded as
\beqa
A&=&\left[1-2e^{-2\chis}+ 2e^{-4\chis} (1-\cot\phis^2 )+\ldots\right]\vare\, ,\nn\\
B &=&\left[1-2e^{-2\chis}-2 e^{-4\chis} (1-\cot\phis^2 )+\ldots\right]\vare^2\, ,\nn
\eeqa
where ellipses contain terms that are higher order in $\exp(-2\chis)$ which may have dependence on $\phis$ too. It is interesting that no dependence on $\phis$ appears in the leading (and even next-to leading) order of these parameters when $\chis\gg1$. This is the limit that we we will focus on in the rest of our analysis. As shown in \cite{Ashoorioon:2013eia}, in this limit maximum separation between the scale of new physics and inflationary Hubble parameter, which is required to utilize the effective field theory, could be obtained.  Also in this limit, it turns out that effectively $\phis\simeq {\pi\over2}$. In this limit, 
\be\label{B-A-relation}
B\approx A^2\,.
\ee
One should note that with decreasing $\chis$, the parameters $A$ and $B$ decrease too. Also the relation between these two parameters, eq. \eqref{B-A-relation}, will not hold any more.

The Planck data indicates $A=0.072\pm 0.022$ on large angular scales with the best fit for the anisotropy direction to be $(l,b)=(227,-27)$. This observational constraint determines $\vare\simeq 0.07\pm 0.022$ and the $\hat n$ direction to be the unit vector along the anisotropy direction. For smaller values of $\chis$, the factor $\gamma_1/\gamma_0<2$, and thus one has to increase the required amount of $\vare$ to account for the observed hemispherical asymmetry. The minimum value for $\vare$, giving rise to the observed hemispherical asymmetry, is hence $\simeq 0.07$. We conclude that 
\be
\vare\gtrsim 0.07\pm 0.022\,.
\ee
In order to induce the asymmetric effect on a finite range of scales, one has to assume that $\vare$ is scale-dependent. This in particular can be realized assuming that the asymmetric effects become more effective when inflaton passes through the scales that left the horizon at very large scales. For example if one assumes that the interaction between the inflaton and the term that induces the asymmetry in the Lagrangian is proportional to $|\phi-\phi_0|^n$, where $\phi_0$ is the value of the inflaton when the scales corresponding to our horizon scale left the horizon. As the inflaton moves away $\phi_0$ creating asymmetric excited quanta becomes more expensive and the effect fades away at smaller scales. 

The existence of quadrupole term, $B (\hat{x}\cdot\hat{n})^2$ with $B\simeq A^2$, besides the dipole term, is one of the predictions of the model. With the observed value of $A$, the coefficient of the quadrupole term turns out to be quite small, $0.0025\leq B  \leq 0.008$. Such would not be discernible from the systematics and noise from the current CMB data. Higher order multipole coefficients are also present in the primordial spectrum, but suppressed by the corresponding power of $\epsilon$. This is a typical feature also among various previously proposed power asymmetry -generating mechanisms, that the predicted couplings of the CMB angular modes falls of with their multipole separation. However, for example, in the non-Gaussianity -generated cases of Ref. \cite{adhikari}, the suppression seems to be squareroot rather than linear $\epsilon^\ell/2$, less sharp than in our case. Higher multipoles can thus be used as a cross-check to distinguish between different explanations for the origin of the dipole asymmetry. 

In the next section we will investigate another signature of the model in the bispectrum, which might be easier to detect.

\section{Bispectrum}

Let us calculate the three-point function for the above direction-dependent excited states to see how they modify the bispectrum. One can calculate the Wightman function for the solution,  $u_k(\tau)$,
 \begin{equation}\label{Wightman}
 G_{k}^{>}(\tau,\tau')\equiv\frac{H^2}{{\dot{\phi}}^2} \frac{u_k(\tau)}{a(\tau)}\frac{u_k^{\ast}(\tau')}{a(\tau')}\,.
 \end{equation}
 The three-point function could be derived from the Wightman function through the following integral \cite{Maldacena:2002vr}:
 \begin{eqnarray}\label{zeta3}
&& \langle \zeta_{\vec k_1}\zeta_{\vec k_2}\zeta_{\vec k_3}\rangle= -i (2\pi)^3 \delta^3\left(\sum \vec k_i \right) \left(\frac{\dot{\phi}}{H}\right)^4 M_P^{-2} H \int_{\tau_{0}}^0 {\rm d}\tau \nonumber\\ && \frac{1}{k_3^2} (a(\tau) \partial_\tau G_{k_1}^{>}(0,\tau)) (a(\tau) \partial_\tau G_{ k_2}^{>}(0,\tau)) (a(\tau) \partial_\tau G_{ k_3}^{>}(0,\tau))\nonumber \\&&+{\rm permutations}+{\rm c.c.}\,,
 \end{eqnarray}
where $\tau_0$ is the moment at which the physical momentum becomes equal to the physical cutoff, $\tau_0\equiv \frac{M}{Hk}$. The Wightman function in the integrand is 
\begin{equation}\label{aGprime} 
\partial_{\tau}G_{k}^{>}(0,\tau))=\frac{H^3}{2\dot{\phi}^2 k}\frac{2(\A-\B)(-\AS~e^{i k\tau}+\BS~e^{-ik\tau})}{a(\tau)}\,.
\end{equation}
The bispectrum takes the form 
\beqa
 &&\langle \zeta_{\vec k_1}\zeta_{\vec k_2}\zeta_{\vec k_3}\rangle=(2\pi)^3 \delta^3 (\sum_{i=1}^{3} \vec k_i ) \frac{2 H^6 \left(\sum\limits_{i>j} k_i^2 k_j^2 \right) }{\dot{\phi}^2 M_P^{2}\prod\limits_{i=1}^3 (2k_i^3 )}\times \nn &&\left[\mathscr{A} \frac{1-\cos(k_t \tau_0)}{k_t} + \mathscr{B} \frac{\sin(k_t\tau_0)}{k_t}+\sum\limits_{j=1}^{3} \mathscr{C}_j \frac{1-\cos(\tilde{k}_j\tau_0)}{\tilde{k}_j}\right. \nn  &&\left. + \sum\limits_{j=1}^{3} \mathscr{D}_j \frac{\sin(\tilde{k}_j\tau_0)}{\tilde{k}_j} \right]\,,
\eeqa
where $k_t=k_1+k_2+k_3$ and $\tilde{k}_j=k_t-2 k_j$. Terms proportional to $\mathscr{C}_j$  and  $\mathscr{D}_j$ are respectively the ones that can lead to enhancement in the local configuration, $k_1\simeq k_2\gg k_3$ \cite{Agullo:2010ws}, or flattened (folded) configuration, $k_1+k_2\simeq k_3$ \cite{Chen:2006nt}. Coefficients, $\mathscr{A, ~B,~C_j}$ and $\mathscr{D}_j$ are as follows
\beqa
\mathscr{A}=\prod (\Ai-\Bi) (\prod \Ais+\prod\Bis)+{\rm c.c.}\nn
\mathscr{B}=i\prod (\Ai-\Bi) (-\prod \Ais+\prod\Bis)+{\rm c.c.}\nn
\mathscr{C}_j=-\prod (\Ai-\Bi) (\frac{\Bjs}{\Ajs} \prod \Ais +\frac{\Ajs}{\Bjs} \prod\Bis  )+{\rm c.c.}\nn
\mathscr{D}_j=i\prod (\Ai-\Bi) (\frac{\Bjs}{\Ajs} \prod \Ais -\frac{\Ajs}{\Bjs} \prod\Bis  )+{\rm c.c.}\nn
\eeqa
\begin{figure}[t]
\includegraphics[angle=0, scale=0.33]{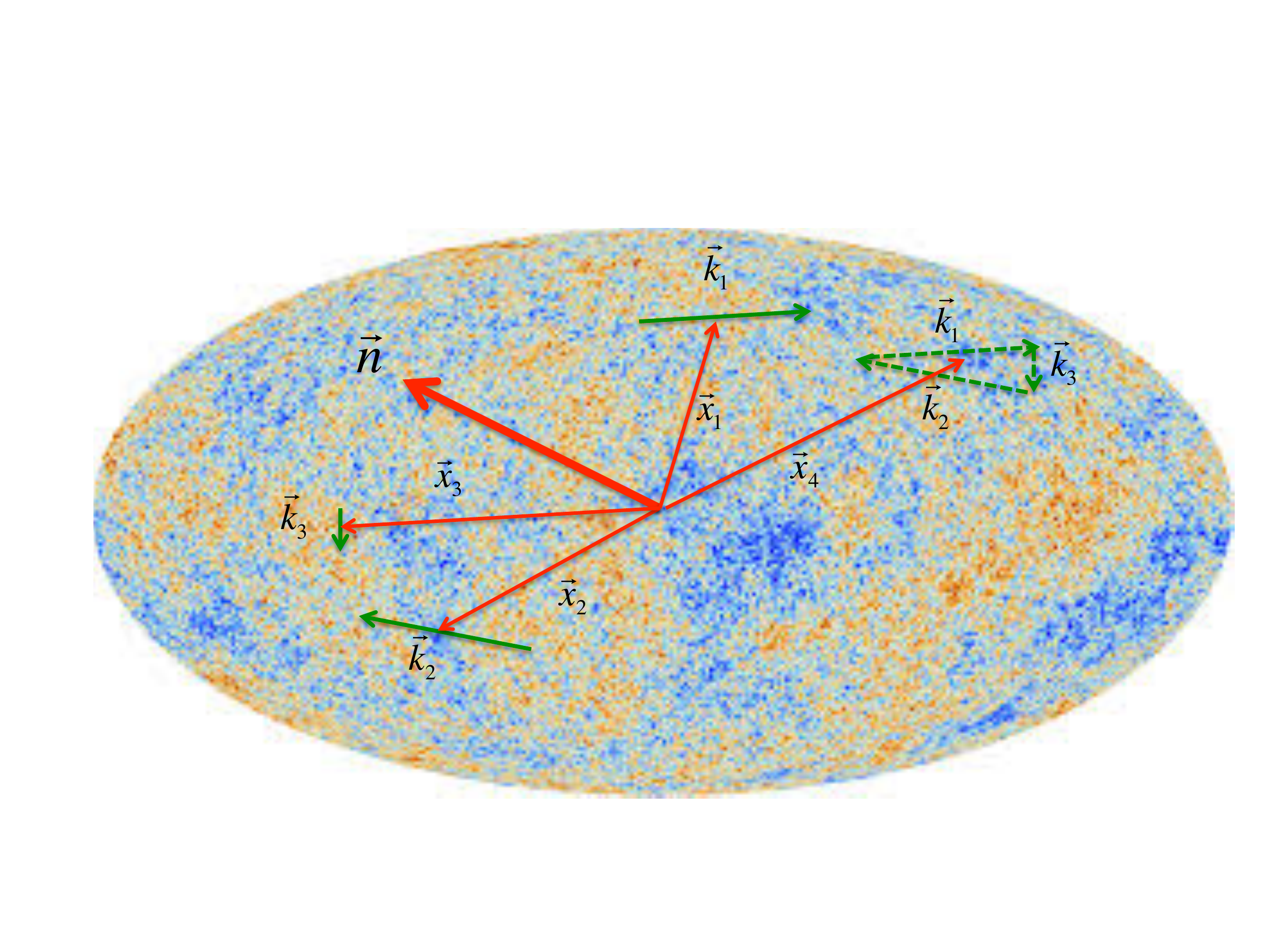}
\caption{The amplitude of non-gaussianity depends on which corner of the sky the mode is located. In a scale-dependent asymmetry, it also depends on whether the asymmetric effects are effective at the scale of interest.}
\label{general-configuration}
\end{figure}

The enhancement of the flattened configuration is however lost in slow-roll inflation after the projection of the bispectrum shape on the 2-dimensional CMB surface \cite{Holman:2007na}. Besides for the large deviations from the Bunch-Davies vacuum where, $\chis\gg 1$ and $\phi\simeq \pi/2$, the enhancement factor is exactly equal to zero.Thus we focus on the local configuration enhancement. Noting that $k_1 \simeq k_2\gg k_3$, the enhancement for the local configuration  three-point function one obtains:
\beqa
 && \langle \zeta_{\vec k_1}\zeta_{\vec k_2}\zeta_{\vec k_3}\rangle\simeq -16 (2\pi)^3 \delta^3 (\sum_{i=1}^{3} \vec k_i )\frac{H^8 \epsilon}{\dot{\phi}^2 \prod\limits_{i=1}^3 (2k_i^3 )} \frac{\sum\limits_{i>j}^{} k_i^2 k_j^2}{k_3}\mathcal{C}\,,\nn
 \eeqa
 where 
 \beqa
\mathcal{C}&=& \mathrm{Re}\left[\prod\limits_{i=1}^3 (\Ai-\Bi) \left(\prod\limits_{i=1}\Ais\left(\frac{\beta_{k_1}^*}{\alpha_{k_1}^*}+\frac{\beta_{ k_2}^*}{\alpha_{ k_2}^*}\right) \right.\right.\nn&&\left.\left.+\prod\limits_{i=1} \Bis\left(\frac{\alpha_{ k_1}^*}{\beta_{ k_1}^*}+\frac{\alpha_{ k_2}^*}{\beta_{k_2}^*}\right)\right) \right]\nn
&=&\mathrm{Re}\left[\prod\limits_{i=1}^3 (\Ai-\Bi)(\alpha_{k_3}^*+\beta_{ k_3}^*) (\alpha_{ k_2}^*\beta_{\vec k_1}^*+\alpha_{k_1}^*\beta_{k_2}^*)\right]\,.\nn
  \eeqa
 
One can calculate the  $ f_{NL}$ parameter using the definition,
\be
 f_{\rm NL}\equiv-\frac{5}{6}\frac{\delta \langle \zeta_{ k_1}\zeta_{ k_2}\zeta_{ k_3}\rangle}{\sum\limits_{i>j}  \langle \zeta_{ k_i}\zeta_{ k_i}\rangle \langle \zeta_{ k_j}\zeta_{k_j}\rangle}\,,
\ee
to be
\be
f_{\rm NL}\simeq-\frac{20 \epsilon}{3}\frac{k_1}{k_3}\frac{\mathcal{C}}{\gamma_S( k_3) (\gamma_S(k_1)+\gamma_S(k_2))}\,.
\ee
Expanding the $ \delta f_{\rm NL}$ in terms of $\vare$ up to second order,
\be
 f_{\rm NL}\simeq f_{\rm NL}^{(0)}+ f_{\rm NL}^{(1)}\vare+ f_{\rm NL}^{(2)}\vare^2\,,
\ee
in the limit that $\chis\gg1$ and thus $\phis\simeq \pi/2$, we have
\beqa
 f_{\rm NL}^{(0)} &\simeq& \frac{5\epsilon}{3} \frac{k_1}{k_3},\\
 f_{\rm NL}^{(1)} &\simeq& \frac{5\epsilon}{3} \frac{k_1}{k_3}\left[\cos(\psi_{\hat x_1})+\cos(\psi_{\hat x_2})-2\cos(\psi_{\hat x_3})\right],\\
 f_{\rm NL}^{(2)} &\simeq& -\frac{5\epsilon}{6} \frac{k_1}{k_3}\left[\cos(\psi_{\hat x_1})^2+\cos(\psi_{\hat x_2})^2-6\cos(\psi_{\hat x_3})^2 \right.\nn&& \left. - 4 \cos(\psi_{\hat x_1}) \cos(\psi_{\hat x_2})+4  \cos(\psi_{\hat x_1}) \cos(\psi_{\hat x_3}) \right.\nn&&+\left. 4  \cos(\psi_{\hat x_2}) \cos(\psi_{\hat x_3}) \right]\,.
\eeqa
 \begin{figure}[t]
\includegraphics[angle=0, scale=0.33]{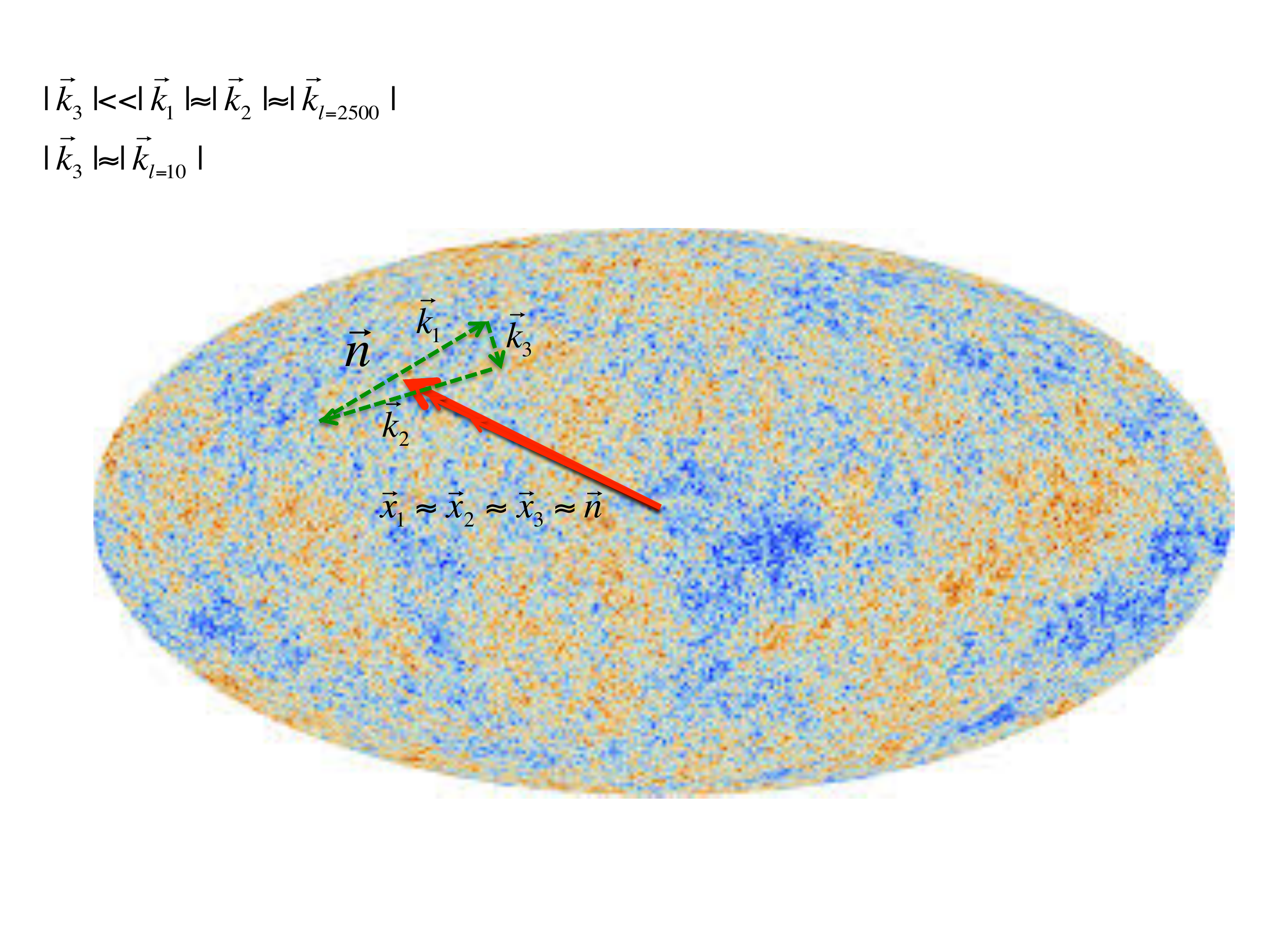}
\caption{The amplitude of local non-gaussianity reaches its minimum departure from the mean value of non-gaussianity, when the considered modes are around the preferred pole of the sky and $|\vec k_3|\approx |\vec k_{l\approx10}|\ll |\vec k_1|\approx |\vec k_2|\approx |\vec k_{l\approx2500}|$. This is because for $\vare( k_1)=\vare(k_2)=0$ and $\vare( k_3)=\vare\neq 0$.}
\label{}
\end{figure}
Hence the amplitude of the bispectrum depends on the angles that three position vectors make with the preferred direction, please see fig.\ref{general-configuration}.

The first term, $ f_{\rm NL}^{(0)}$, which gives the dominant contribution to the bispectrum, is independent of the these angles though. For an inflationary model with $\epsilon\approx 0.01$, the 
\be\label{meanfnl}
 f_{\rm NL}^{(0)}\simeq 4.17\,,
\ee
where we have taken the largest scale at which the cosmic variance is negligible to be corresponding to $l_{\rm min}\simeq10$ and the smallest one to be the largest $l$  probed by the Planck experiment, $l_{\rm max}\simeq 2500$ \footnote{There is a minor correction to to the value of $f_{\rm NL}$ as quoted in \cite{Ashoorioon:2013eia} due to a missing factor of $\approx 10$ in the previous analysis.}. This is within the $2\sigma$ allowed region of local nongaussianity from the Planck 2015 experiment \cite{Ade:2015ava}.
 \begin{figure}[t]
\includegraphics[angle=0, scale=0.33]{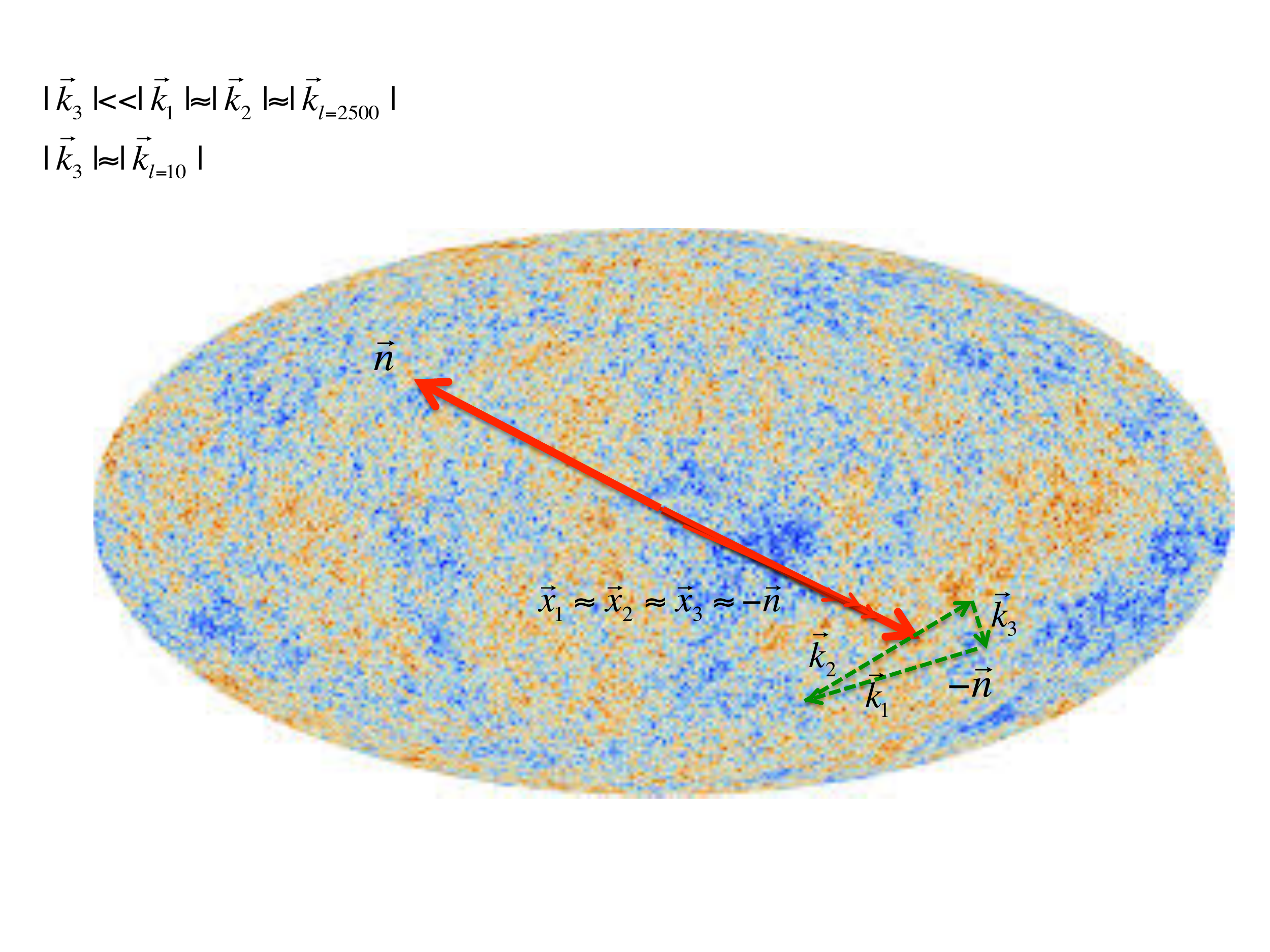}
\caption{The amplitude of local non-gaussianity reaches its maximum departure from the mean value of non-gaussianity, when the considered modes are around the antipode of the preferred pole and $|\vec k_3|\approx |\vec k_{l\approx10}|\ll |\vec k_1|\approx |\vec k_2|\approx |\vec k_{l\approx2500}|$.}
\label{}
\end{figure}
At higher orders, the excited asymmetric initial condition induces directional dependence to the bispectrum at the first order correction. It is easy to verify that if $\hat x_1=\hat x_2=\hat x_3$ and asymmetry is scale-independent, the directional dependence vanishes at both first and second order in $\vare$. That is if the chosen momenta are all at the same corner of the sky, there is no modulation on top of the mean value \eqref{meanfnl}. On the other hand if the three modes are at different corners of the CMB sky, even if the asymmetry is scale-independent, one will see modulation on top of the mean non-gaussianity value \eqref{meanfnl}. The maximum of the modulation occurs when the short wavelength modes, $k_1$ and $k_2$, are at the preferred pole, $\cos(\psi_{\hat x_1})=\cos(\psi_{\hat x_2})=1$ and the  long wavelength mode is at its antipode. For $\vare=0.07$, the maximum non-gaussianity obtained at the scales is 
\be
 f_{\rm NL}^{\rm max}\simeq f_{NL}^{(0)}(1+4\vare+10\vare^2)\approx 5.54\,.
\ee
On the other hand, the minimum shift from the Bunch-Davies value result, corresponding to minimum value for non-gaussianity, would occur for when the short wavelength modes are the antipode pole and the long wavelength mode is at the preferred pole. For this configuration
\be
 f_{\rm NL}^{\rm min}\simeq f_{NL}^{(0)}(1-4\vare+10\vare^2)\approx 3.20\,.
\ee

Since the hemispherical asymmetry parameter is scale dependent and  fades aways at $l\gtrsim600$, one should reconsider the above result. In fact the first order correction to the mean value of non-gaussianity, which is the dominant modulation term, could be written as
\beqa
&&\Delta f_{\rm NL}^{(1)} =\nn
&&\frac{5\epsilon}{3} \frac{k_1}{k_3}\left[\vare(k_1)\cos(\psi_{\hat x_1})+\vare(k_2)\cos(\psi_{\hat x_2})-2\vare(k_3)\cos(\psi_{\hat x_3})\right]\,,\nn
\eeqa
maximum modification from the mean value \eqref{meanfnl}, occurs for the configuration where $k_1\simeq k_2$ is the wavenumber corresponding to $l\approx 2500$ and $k_3$ corresponds to $l=10$. In this case, $\vare(k_1)=\vare(k_2)=0$ and $\vare(k_3)=\vare$. Now if the large wavelength mode, $k_3$ is around the preferred pole, $\cos(\psi_{\hat x_3})=1$, we will have a decrement from the mean value of non-gaussianity 
\be
 f_{\rm NL}^{\rm min}\simeq f_{NL}^{(0)}(1-2\vare+3\vare^2)\approx 3.64\,.
\ee
It does not matter at which corner of the sky, $k_1$ and $k_2$ modes are located. We can assume that they are centered around the preferred pole too.

On the other hand, If $\vec k_3$ is around the antipode of the preferred pole, there will be an enhancement in the mean value of local nongaussianity
\be
 f_{\rm NL}^{\rm max}\simeq f_{NL}^{(0)}(1+2\vare+3\vare^2)\approx 4.81\,.
\ee
The difference between these two values of non-gaussianity at two poles, $\delta f_{\rm NL}\approx 1.17$, can be used to distinguish this scenario from the competing proposals that try to explain the observed hemispherical asymmetry.

We recall non-gaussianity can increase the probability of a power asymmetry. The impact of higher order nongaussinity has been also considered \cite{adhikari,Kenton:2015jga}. A high $g_{NL}$, even if occurring in an isotropic vacuum, could induce anisotropic ''power asymmetry'' on the $f_{NL}$. It is an open question at the moment, whether a $g_{NL}$, while staying within the observational bounds, could induce anisotropies in the $f_{NL}$ at the order we are predicting.

 \section{Conclusion}
  Hemispherical asymmetry, is one of the persistent forms of isotropy violation that has been observed in both  WMAP and Planck data. The hemispherical asymmetry observed is scale-dependent and vanishes for $l\gtrsim 600$. In this paper we suggested a scenario that accounts for the observed hemispherical asymmetry using asymmetric initial condition. We noticed that there are infinite higher multipole corrections to the power spectrum. In the limit where the scale of new physics is maximally separated from the Hubble parameter, only the dipole and quadrupole terms survive. Requiring that the amount of asymmetry is as large as the observed value, we obtained a small but finite amplitude of quadrupole correction to the power spectrum. This is one way to distinguish this scenario from other scenarios that try to explain the origin of hemispherical asymmetry. We showed that the model exhibits unique signatures in the bispectrum. Due to the excited initial states, the local configuration gets enhanced around a mean value which for a large field inflationary model, $\epsilon \approx 0.01$, turns out to be around $4.17$. There will be modulations on top of this mean value which is dependent upon the angle that the patches that contain the mode make with the preferred direction. The amount of variation minimizes for the configuration in which the short and long wavelengths modes are  around the preferred pole and $|\vec k_3|\approx |\vec k_{l\approx10}|\ll |\vec k_1|\approx |\vec k_2|\approx |\vec k_{l\approx2500}|$ with $f_{\rm NL}^{\rm max}\approx 4.81$. The maximum occurs when these modes are at the antipode of the preferred pole, $f_{\rm NL}^{\rm min}\approx 3.64$. The difference of non-gaussianity between these two configurations is as large as $\simeq 1.17$ which would be a definite indication of the model.
 \acknowledgments
We specially would like to thank Zac Kenton for commenting on the first version of the article.  We are also thankful to Y. Akrami and H. Firouzjahi for illuminating discussions, and to E. Komatsu, D. Lyth, D. Spergel and M. M. Sheikh-Jabbari for very helpful comments on the draft.

\end{document}